\documentclass[aps,prb,showpacs,twocolumn,superscriptaddress]{revtex4}
\usepackage{mathrsfs}
\usepackage{epsfig}
\usepackage{graphicx}
\usepackage{amsfonts}
\usepackage[figuresright]{rotating}
\usepackage{amssymb}
\usepackage{amsmath}
\usepackage{dcolumn}
\usepackage{bm}
\usepackage{color}

\def\be{\begin{equation}} \def\ee{\end{equation}}
\def\bea{\begin{eqnarray}} \def\eea{\end{eqnarray}}

\newcommand{\lmu} {Department of Physics and Arnold Sommerfeld Center for Theoretical Physics,
Ludwig-Maximilians-Universit{\"a}t M{\"u}nchen, Theresienstr.\ 37,
80333 Munich, Germany}
\newcommand{\sjtu} {Key Laboratory of Artificial Structures and Quantum
Control, Department of Physics and Astronomy, Shanghai Jiao Tong University, Shanghai 200240, People's Republic of China}

\newcommand{\fudu} {Department of Physics and State Key Laboratory of Surface Physics, Fudan University, Shanghai 200433, China}

\newcommand{\CICAM} {Collaborative Innovation Center of Advanced Microstructures, Nanjing 210093, China}

\newcommand{\wqc} {Wilczek Quantum Center, Shanghai Jiao Tong University, Shanghai 200240, China}

\begin{document}
%\title{Dissipative quantum many-body systems: a  Quantum Monte Carlo study}
\title{Interacting lattice systems with quantum dissipation: a quantum Monte Carlo study}

\author{Zheng Yan}
\affiliation{\fudu}
\affiliation{\CICAM}

\author{Lode Pollet}
\affiliation{\lmu}
\affiliation{\wqc}

\author{Jie Lou}
\affiliation{\fudu}
\affiliation{\CICAM}

\author{Xiaoqun Wang}
\affiliation{\sjtu}
\affiliation{\CICAM}

\author{Yan Chen}
\email{yanchen99@fudan.edu.cn}
\affiliation{\fudu}
\affiliation{\CICAM}

\author{Zi Cai}
\email{zcai@sjtu.edu.cn}
\affiliation{\sjtu}

\begin{abstract}
Quantum dissipation arises when a large system can be split in a quantum system and an environment  where the energy of the former flows to. Understanding the effect of dissipation on quantum many-body systems is of particular importance due to its potential relations with quantum information. We propose a conceptually simple approach to introduce the dissipation into interacting quantum systems in a thermodynamical context, in which every site of a 1d lattice is coupled off-diagonally to its own bath. The interplay between quantum dissipation and interactions gives rise to counterintuitive interpretations such as a compressible zero-temperature state with spontaneous discrete symmetry breaking and  a thermal phase transition in a one-dimensional(1D) dissipative quantum many-body system as revealed by  Quantum Monte Carlo path integral simulations.
%, which enable us to implement the Quantum Monte Carlo method to investigate the properties of the dissipative quantum many-body systems. The interplay between the dissipations and interactions gives rise to novel phases and phase transitions, including a compressible zero-temperature state with spontaneous discrete symmetry breaking, and a finite temperature phase transitions in one-dimensional dissipative quantum many-body system.
\end{abstract}

%\pacs{67.85.Pq, 67.60.Fp, 05.30.Rt}
%\pacs{05.30.Jp, 75.10.Pq, 02.70.Ss, 03.65. Yz}
\pacs{75.10.Pq, 02.70.Ss, 05.30.Jp,  03.65. Yz}
\maketitle

\section {Introduction}
Almost all the quantum systems are inevitably coupled to their surroundings.  Understanding a quantum system immersed in an environment is not only of immense practical significance in quantum simulation and information processing~\cite{Schlosshauer2007}, but also in understanding fundamental questions such as the quantum measurement process, quantum-to-classic crossover/transitions, and the mechanism of decoherence~\cite{Hanggi2006,Hanggi2008,Caldeira1981, Caldeira1983,Caldeira1983b,Leggett1987,Weiss1999,Breuer2002,Gardiner1999,Leggett2004}.  The  scenario is further complicated if the open quantum system itself is a (strongly) interacting many-body system, which is indeed the case for the majority of current quantum computing systems including trapped ions, Rydberg atoms,  and solid-state quantum computers/simulators.  In such systems,  the interplay between the quantum many-body effects and the dissipation gives rise to a plethora of novel phenomena~\cite{Chakravarty1986,Fisher1987,Cazalilla2006,Torre2010,Zhu2015,Werner2004,Werner2005b,Sperstad2011,Stiansen2012,Cai2014,Diehl2008,Verstraete2009,Prosen2008,Daley2009,Wilming2017,Rancon2013,Schindler2013,Johnson2011} in fields as  diverse as solid state physics,  quantum information and atomic physics.

%Statistical physics textbooks typically assume that a system, coupled to a large heat bath capable of absorbing the  energy of the system, is irreversibly driven towards thermodynamic equilibrium independent of the system-bath (SB) coupling strength. This property, true for equilibrium classical systems, is in general no longer valid for quantum systems -- especially those beyond the weak SB coupling limit. In such cases the quantum system entangles with the bath turning its reduced density matrix explicitly dependent on the SB coupling strength. The quantum system can then no longer be described by the  Boltzman-Gibbs ensemble~\cite{Hanggi2006,Hanggi2008}. In a dissipative quantum system beyond the weak SB coupling limit, the competition between the quantum fluctuations and dissipation gives rise to numerous novel effects that are of particular importance for understanding  fundamental questions such as the quantum measurement  process, quantum-to-classic crossover/transitions, and the mechanism of decoherence~\cite{Caldeira1981, Caldeira1983,Caldeira1983b,Leggett1987,Weiss1999,Breuer2002,Gardiner1999,Leggett2004}.  Furthermore, such systems are also of immense practical significance in quantum simulation and information processing~\cite{Schlosshauer2007}, especially for the rapidly developing field of scalable quantum computing.

A variety of theoretical and numerical methods for open quantum systems has been employed. Most of them follow the spirit of taking the environment  into account in an exact or approximate way through deriving an effective action by integrating out the bath degrees of freedom~\cite{Caldeira1981, Caldeira1983,Caldeira1983b,Leggett1987,Weiss1999,Breuer2002,Gardiner1999}. Despite its great success, this treatment is difficult --  or at least impractical in dealing with the complex situations where the system itself is a quantum many-body system. Except for some special cases~\cite{Exceptions}, often in the field of open quantum systems, tracing out the bath degrees of freedom produces an effective action with unequal-time (retarded) interactions in imaginary or real time. This complicates the quantum many-body system usually to such a degree that  numerical methods currently used in strongly correlated physics can no longer be applied~\cite{Exception2}. A different route to study open quantum many-body systems follows the strategy of quantum optics by generalizing the Born$-$Markov master equation to the many-body case, which is restricted to those open quantum systems with a weak system-bath (SB) coupling to a Markovian environment, neglecting the time delay in the interactions.

%Apart from the intrinsic difficulties in strongly correlated physics, a proper modeling and dealing with the environment is also a theoretical challenge considering  the large variety and degrees of freedom of the real environment in nature.
% In general, dissipation arises when a large system can be separated into a subsystem and the remaining, the former is the cynosure in the global system while the other is regarded as bath, which damps the dynamics of the system by dissipating its energy. In spite of its large variety in nature, a bath can be modeled either by a set of oscillators  or spins, representing two universality classes of environment respectively.

In this paper, we propose a conceptually simple approach to study the dissipative interacting quantum system by treating the bath degrees of freedom on the same footing as the system variables, even though we are only interested in the properties of the quantum system. An essential ingredient of this approach is the separation of a global system into a quantum subsystem and the bath, where the choice of the bath Hamiltonian needs to be  simple enough to be  tractable by conventional many-body numerical methods, yet complicated enough to capture the essential physics of such environments occurring in nature.  Motivated by a recent intriguing proposal of modeling the environment by an engineered spin chain~\cite{Ramos2016,Vermersch2016}, we perform a numerically exact Quantum Monte Carlo (QMC) path integral simulation with the worm algorithm~\cite{Prokofev1998} (here in the implementation of Ref.~\cite{Pollet2005}) and the stochastic series expansion algorithm~\cite{Sandvik2002} to study a composed quantum many-body system with a special lattice geometry, which is interpreted as an interacting lattice system with quantum dissipation. \\
\begin{figure}[htb]
\includegraphics[width=0.99\linewidth,bb=1 6 420 181]{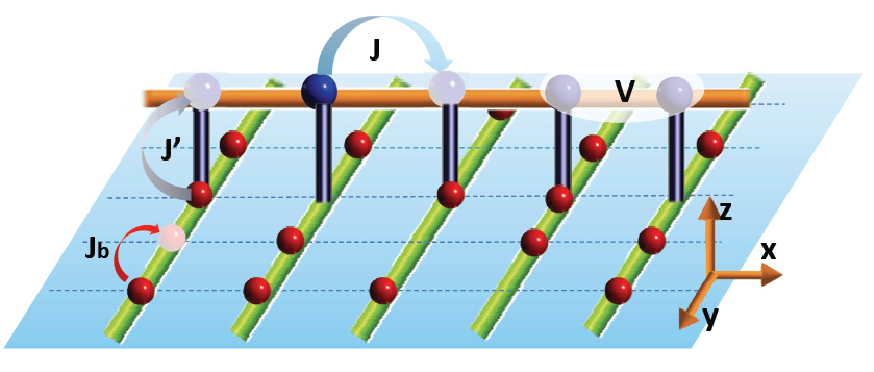}
\caption{ Setup of an interacting lattice model with a local quantum dissipation mechanism: An interacting hard-core boson chain (orange) is off-diagonally and uniformly coupled to independent bath chains (green).  }
\label{fig:dissipation}
\end{figure}

The main points of our results are highlighted as
follows. The phase diagram of an interacting lattice system with quantum dissipation is investigated by a numerically exact method. At zero temperature, it is shown that the dissipation can fundamentally alter the properties of the 1D system, and induce a novel state absent as a ground state in typical 1D closed quantum systems. At finite temperature, we find that even though both the dissipation and the thermal fluctuations are individually detrimental to long-range order,  their conspiracy can facilitate it and give rise to a finite-T phase transition in this 1D dissipative quantum many-body system.  Our results also show that spontaneous symmetry breaking can only occur in a subsystem spatially embedded in a larger system with an inhomogeneous Hamiltonian. \\

\section {Model and method}

 The Hamiltonian describing a dissipative  system contains three parts: $H_{tot}=H_s+H_b+H_{sb}$.  For the system Hamiltonian $H_s$ we choose a prototypical example of an interacting quantum model: a 1D hard-core boson chain with nearest-neighboring density-density interactions (or, equivalently, an XXZ model in the spin language), which reads:
\begin{equation}
H_s=\sum_i -J(a_i^\dag a_{i+1}+h.c)+V (n_i-\frac 12) (n_{i+1}-\frac 12), \label{Eq:XXZ}
\end{equation}
where $a_i^\dag$ ($a_i$) is the creation(annihilation) operator of a hard-core boson on site $i$, $J$ is the tunneling amplitude, and $V>0$ denotes the repulsive interaction strength.\\
 Apart from the intrinsic difficulties in solving strongly correlated physics, a proper modeling and dealing with a realistic environment is also a theoretical challenge. For many realistic quantum systems we do not have a proper understanding of the microscopic origin of dissipation. %Such difficulties  motivate us to choose the  simplest possible model resembling an environment in nature.
 A desired bath model consists of  a quantum system with gapless excitation spectrum and its number of degrees of freedom should be much larger than that of the system it is coupled to. In addition, we devise the baths surrounding different system sites in such a way that they do not influence each other, reflecting the situation in  scalable quantum computing with solid-state devices. Finally, the bath model needs to be as simple as possible to be tractable by numerical methods. A minimal bath model satisfying the above requirements is a set of independent chains of hard-core bosons, each coupled to a system site as shown in Fig.~\ref{fig:dissipation}, with Hamiltonian:
\begin{equation}
H_b=-\sum_{i,j}J_b(b^\dag_{i,j}b_{i,j+1}+h.c)\label{Eq:bath}
\end{equation}
where $b^\dag_{i,j}$ ($b_{i,j}$) denotes the  creation(annihilation) operator of a hard-core boson at site $j$ of the bath linked to the system on site $i$.  However, we should emphasize that the bosons in the bath need not have the meaning of physical particles, nor the 1D structure that of the real geometry of the baths in nature. $H_b$ in Eq.(\ref{Eq:bath}) plays the role of a quantum reservoir with continuum spectrum that can absorb extra energy from the system.
 In passing we note that the idea of independent baths for composite systems was first introduced in laser theory and later in modeling heat conduction~\cite{Davies1978}.  In mesoscopic physics, Buttiker has proposed the idea of using a local reservoir to study the environment-induced inelastic scattering in quantum transport~\cite{Buttiker1985,Buttiker1986}. However, in none of the above cases have interactions  been considered.
 \\
Regarding the SB coupling,  a system can either interact diagonally with the bath preserving its particle number, or off-diagonally, exchanging both energy and particles simultaneously. %, respectively corresponding to the canonical or grand canonical ensembles in the weak SB coupling limit.
The former case has been investigated previously by two of us using a different algorithm~\cite{Cai2014}. We therefore  focus here on the off-diagonal,  particle exchange SB coupling with the Hamiltonian:
\begin{equation}
H_{sb}=-\sum_i J'(a^\dag_ib_{i,0}+h.c).\label{Eq:SB}
\end{equation}
We assume that each system site couples only to the central site $j=0$ of the bath chain with coupling strength $J'$. In spite of their simplicity, Eq.(\ref{Eq:bath}) and Eq.(\ref{Eq:SB})  capture some of the essential physics and represent at least part of the physical reality  in both a quantum optics and solid state context; {\it e.g.}, the spectrum of the bath resembles the one of a photonic band gap material, whereas the SB coupling, after a proper transformation, is reminiscent of the the matter-light interaction in quantum optics.  In addition, both the SB coupling strength and  the bath properties ({\it e.g.} the memory time) are highly tunable, which enables us to investigate both  Markovian and non-Markovian physics ranging from  weak to strong SE coupling  regimes in a unified picture.

The most obvious advantage of using Eq.(\ref{Eq:bath}) and Eq.(\ref{Eq:SB}) to mimic dissipation  is that it enables us to  treat the bath on the same footing as the system.  In the following, we use Quantum Monte Carlo (QMC) simulations to study the composed quantum system in the comb-like geometry, in which the 1D subsystem is regarded as  \textquotedblleft system\textquotedblright and the remainder as \textquotedblleft bath(s)\textquotedblright. Even though we solve the global system,  {\it we focus only on the properties of the system}. In our simulations, we assume that the composed system (system + bath) is weakly coupled to a superbath with working temperature $T$ and is thus in thermal equilibrium even though the system is entangled with the bath.
%However, {\color{red} the system itself is away from thermal equilibrium} since it entangles with the bath. {\color{red}
In all simulations, we focus on  half-filling for the composed system. Without loss of generality, we assume that the system and bath chain have the same length $L$. Periodic boundary conditions are used. Both the ground state and the finite-temperature properties of the composed system are investigated. \\

\begin{figure}[htb]
\includegraphics[width=0.48\linewidth,bb=49 322 487 743]{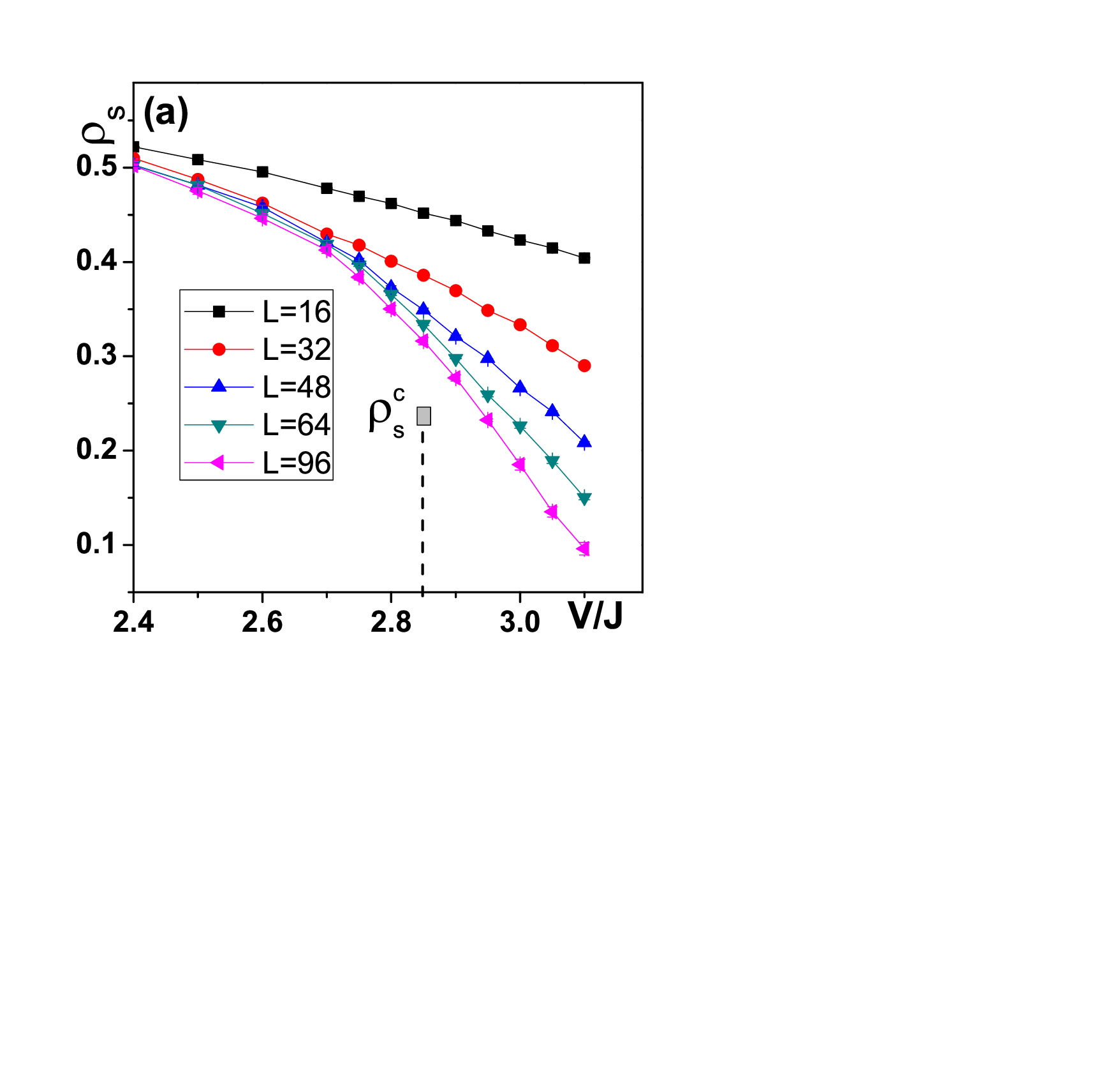}
\includegraphics[width=0.49\linewidth,bb=38 320 488 742]{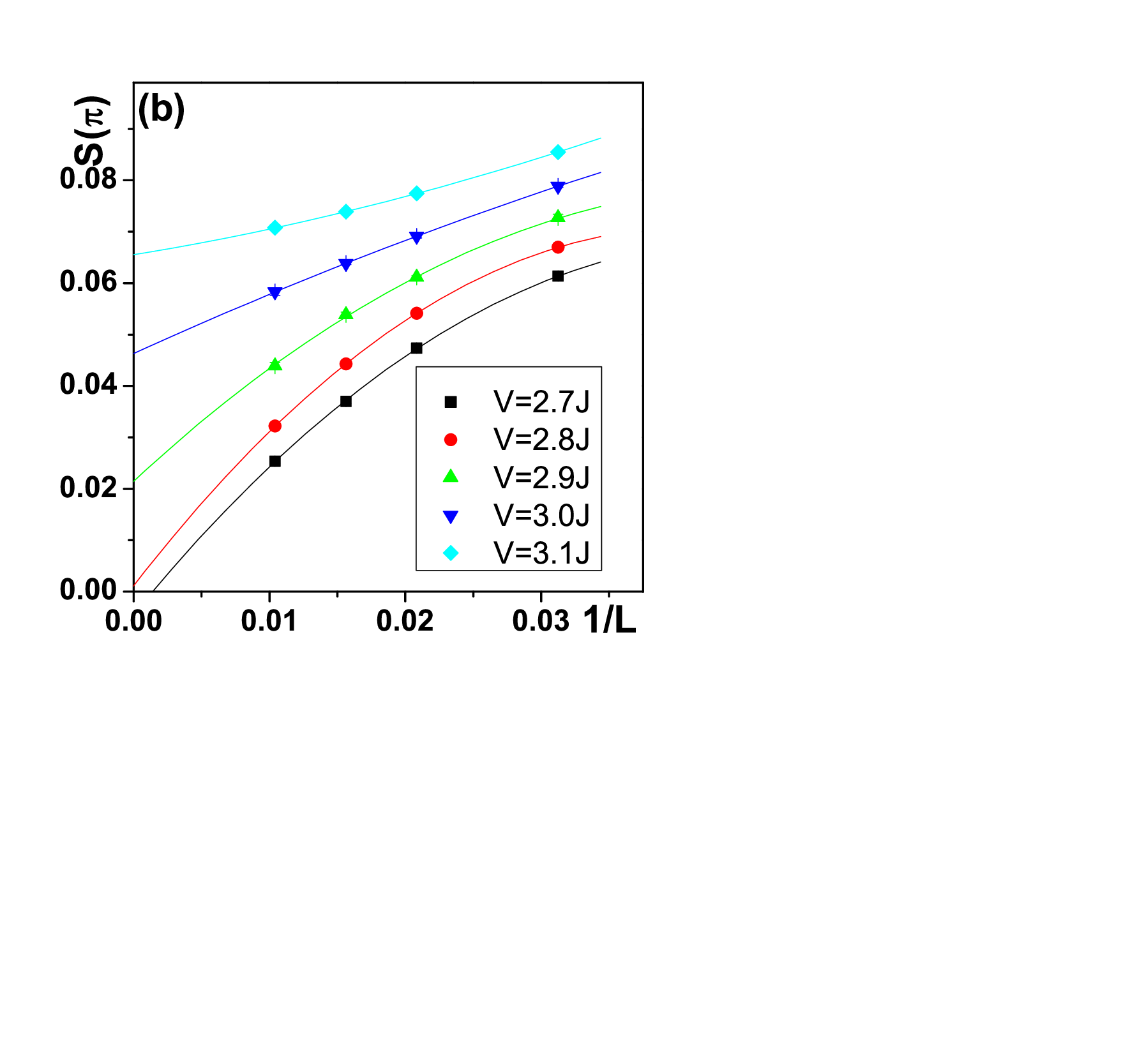}
\includegraphics[width=0.98\linewidth,bb=66 200 488 470]{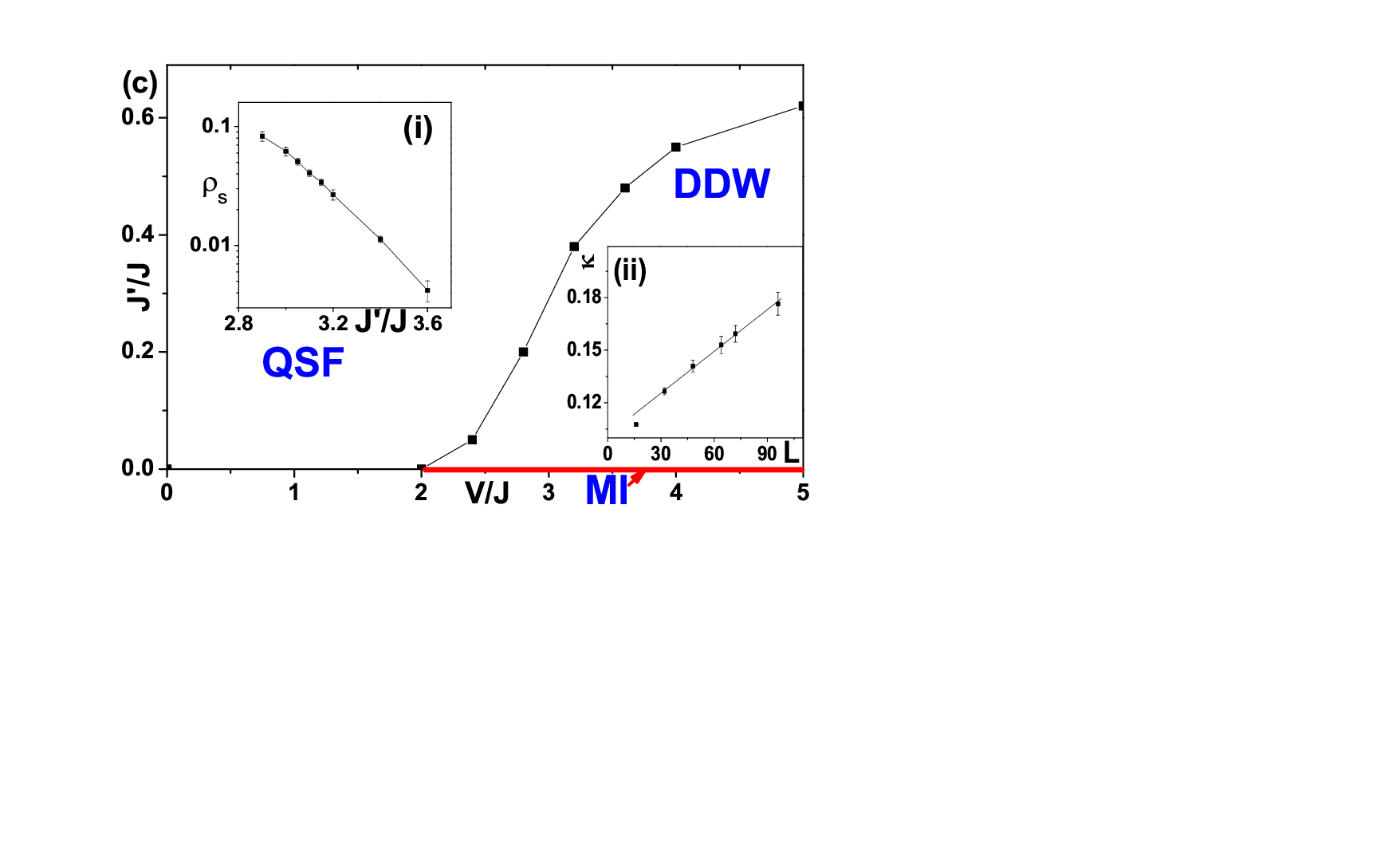}
\caption{(a) The superfluid density of the system as a function of $V$ for various system sizes and $\beta$; (b) the finite size scaling of the DW structure factor for various $V$; [$J'=0.2J$ for (a) and (b)] (c) the zero temperature phase diagram of the system with three different phases: the Luttinger liquid (LL), the Mott-insulating phase (MI) with  CDW order and the dissipative density wave (DDW) state. Inset: (i) $\rho_s$ as a function of $J'$  in the LL phase with  $V=0$; (ii) finite size scaling of $\kappa$ in the compressible DW phase with parameters $V=4J$, $J'=0.2J$. The inverse temperature is scaled as $\beta=L$ in (a)-(c).  }
\label{fig2}
\end{figure}

\section {Phase diagram at zero temperature: an infinitely compressible insulating phase with density modulations}
We first focus on the ground state ($T=0$) of the composed system. In the QMC simulations, we scale the inverse temperature as $\beta=L$, as is relevant for Luttinger liquids, thus setting the dynamical critical exponent $z=1$. The thermodynamic limit is approached in the limit $L\rightarrow \infty$.  Without the SB coupling ($J'=0$) it is known that 1D hard-core bosons with NN interaction at half-filling undergo a Kosterlitz-Thouless (KT)-type quantum phase transition at $V=2J$ from a Luttinger liquid to a gapped Mott insulator with charge-density-wave (CDW-MI) order. We investigate  how the SB coupling changes the above picture. To gain insight, we first discuss qualitatively the effect of the SB coupling on the respective phases.
Since a particle which escapes into the $y$-direction must come back to the same site, the bath can be seen as enlarging the unit cell from a single site to an (infinitely long) chain. Hence, we do not expect the 1d nature of the Luttinger liquid to change substantially, but since the system in the $y$-direction is gapless and compressible the charge per unit cell does not need to be quantized.

%To gain insight, we first provide a qualitative analysis on the effect of the dissipation. In the absence of the SB coupling the hard-core bosons move coherently along the 1D chain  in the Luttinger liquid phase, thus building up quasi-long range off-diagonal coherence in the system. In the presence of the SB coupling, the coherent movement of a boson in the system is interrupted since the boson can escape into the bath, which gives rise to a finite lifetime for the bosonic coherent excitations in the systems.  In the CDW Mott insulating phase, the coupling to a quantum bath introduces extra quantum fluctuations which may be detrimental to both the Mottness and CDW long-range order.

The first quantity we analyze is the superfluid density, defined as $\rho_s=L/\beta \langle W^2\rangle$ with  $W$  the winding number defined along the $x$-direction~\cite{Ceperley1987}. In Fig.2 (a), we fix the SB coupling $J'=0.2J$ and plot $\rho_s$ as a function of $V$. A Weber-Minnhagen fit~\cite{Weber1988} shows that $\rho_s$ exhibits a sudden drop from a finite value $\rho_s^c=0.228$ to zero at $V=2.85J$(see the Suppl. Mat.~\cite{Supplementary}), indicating that there is still a KT-type quantum phase transition. The non-Luttinger liquid phase on the other side of the transitions is also characterized by spontaneous translational symmetry breaking through the emergence of density-wave (DW) order characterized by the static DW structure factor, $S(\pi)=1/L^2 \sum_{ij}(-1)^{i-j}\langle(n_i-\frac 12)(n_j-\frac 12)\rangle$,  as shown in Fig 2 (b). Up to now, it seems that the dissipation doesn't bring any qualitative change to the system except shifting the position of the phase transition point. A striking difference can however  be found in the compressibility of the DW phase, defined as $\kappa=\beta/L (\langle N^2\rangle-\langle N\rangle^2)$  with $N$ the total particle number within the system chain (not the total system).
As shown in Inset (ii) of Fig. 2 (c),  the compressibility $\kappa$  increases linearly with $L$ in the dissipative density wave (DDW) phase, indicating an infinitely compressible state in the thermodynamic limit which makes it fundamentally different from the CDW Mott insulating state at $J'=0$. The divergence of the compressibility  is due to the fact that every system site is coupled to an infinite number of degrees of freedom; thus, every unit cell can be doped with no energy cost. In passing, we note that anomalously large isochoric compressibilities have in the past been found in totally different contexts such as the superclimb of edge dislocations in solid $^4$He~\cite{Ray2008,Soyler2009}.
Furthermore, the vanishing of superfluid density in the direction of the DW modulation rules out a supersolid. Such a compressible insulating phase is absent as a ground state in typical closed quantum  spin-systems.
The SB coupling allows particles to delocalize over a larger system and is expected to enhance the superfluid phase. Indeed, when increasing $J'$ at fixed $V > 2J$, the system goes over from the CDW-MI to the DDW phase and finally to a superfluid, as can be seen in Fig.~\ref{fig2}, panel (c). However, further increasing $J' \gg J$ (which no longer corresponds to a physically motivated SB coupling) enhances  singlet formation and leads to a strong suppression of $\rho_s$ , as can be seen in inset (i), but we expect no phase transition in the thermodynamic limit.

\begin{figure*}[htb]
\includegraphics[width=0.32\linewidth,bb=116 53 755 538]{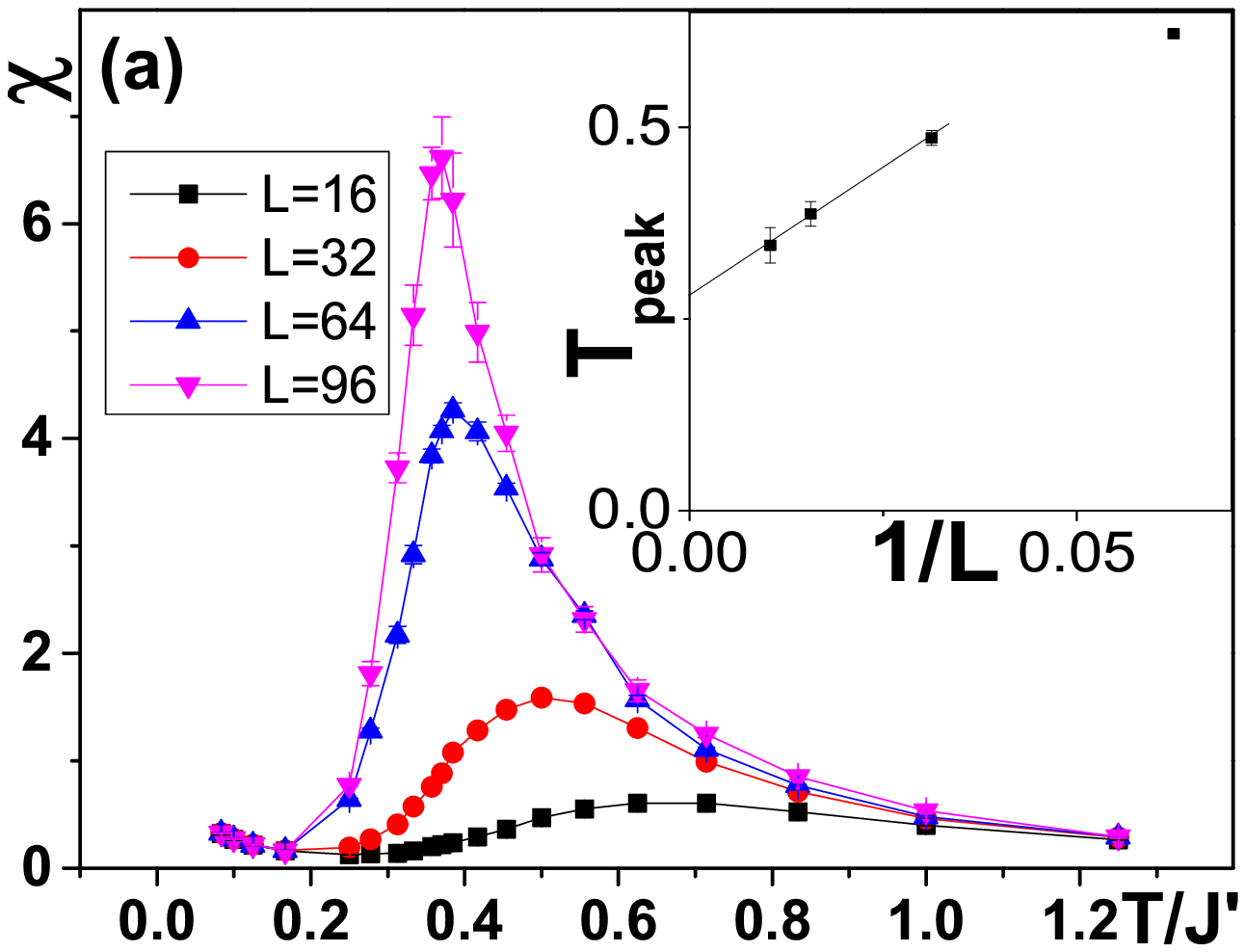}
\includegraphics[width=0.335\linewidth,bb=66 58 735 531]{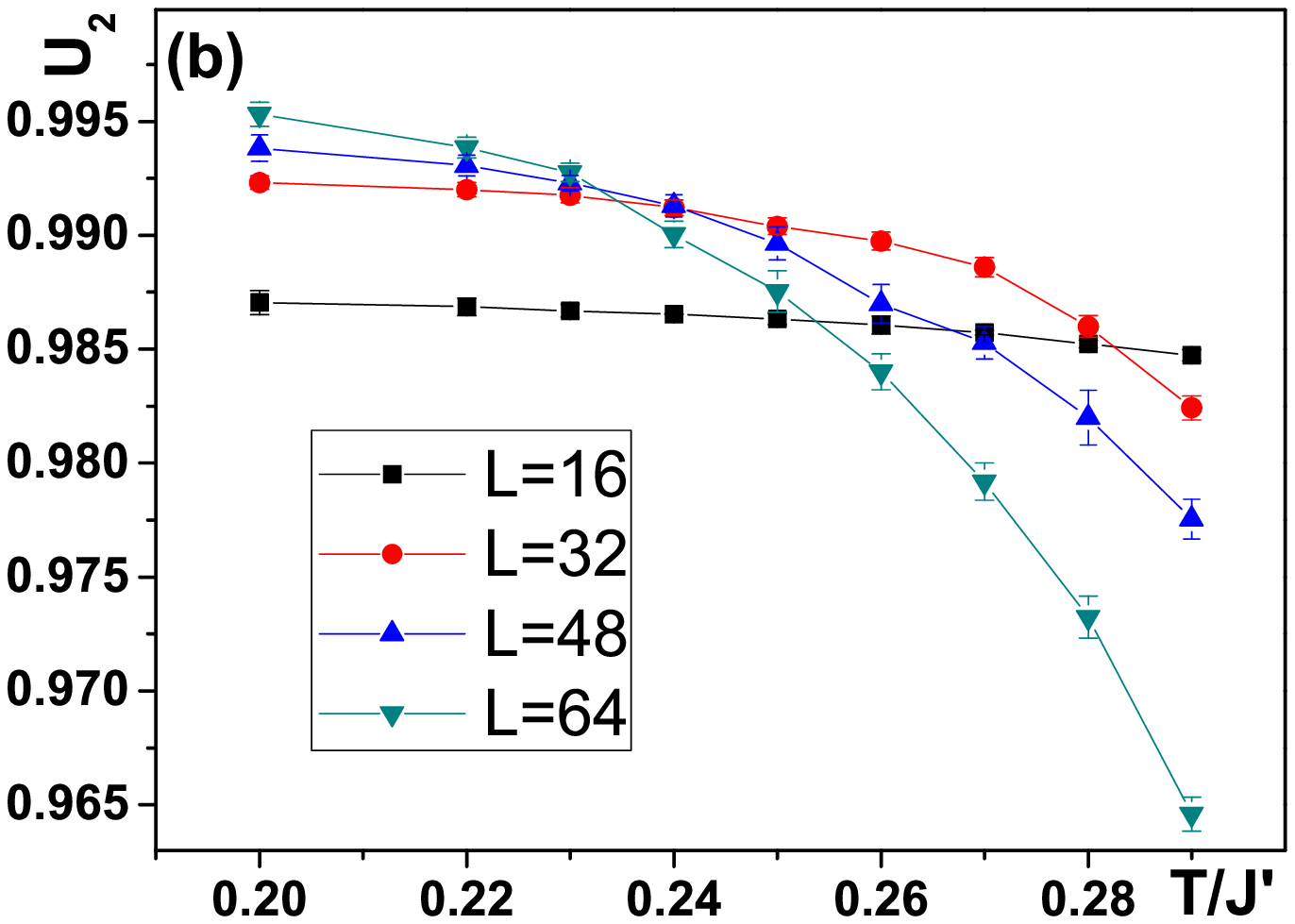}
\includegraphics[width=0.32\linewidth,bb=99 60 734 533]{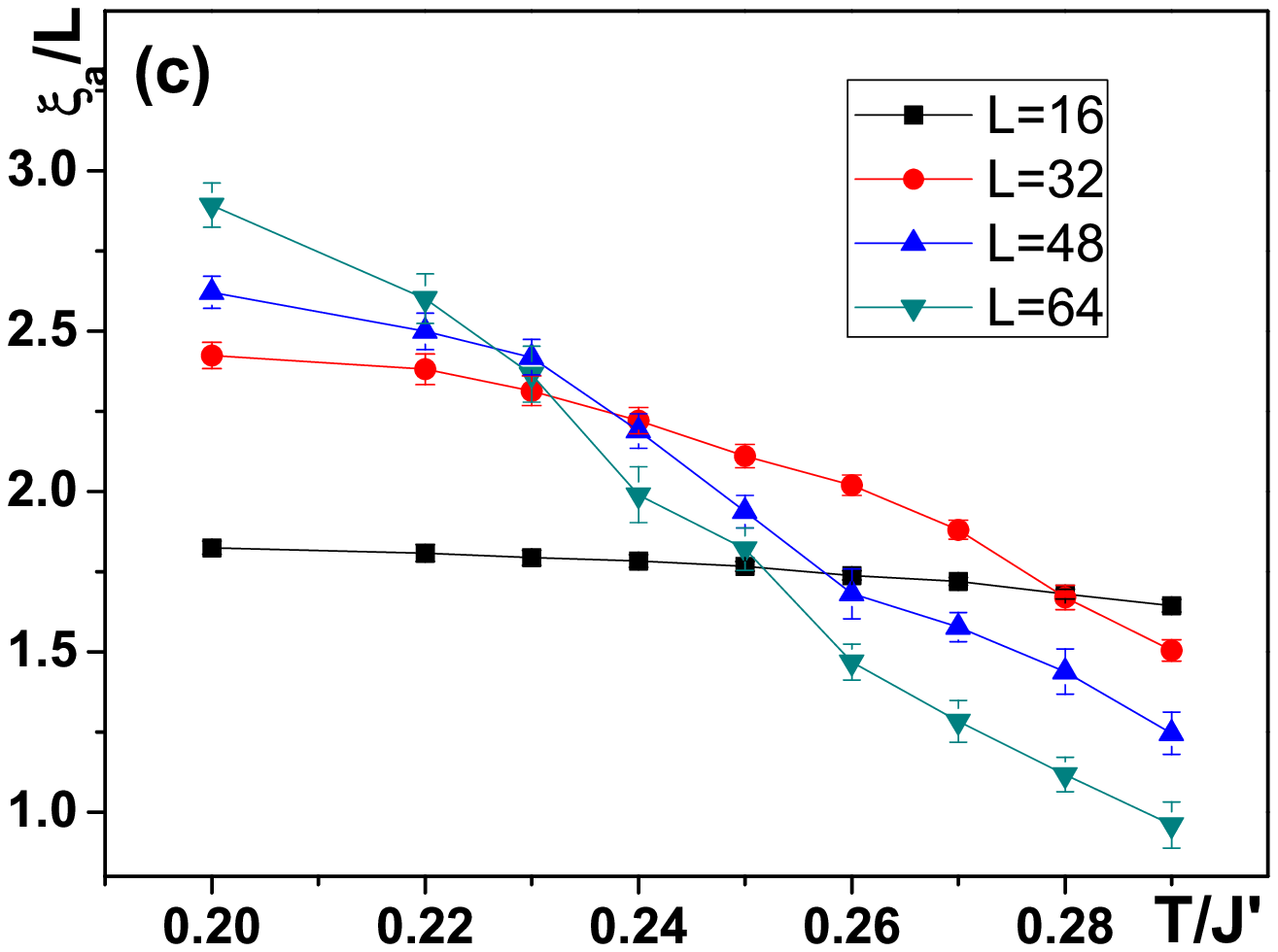}
\caption{(a) The staggered susceptibility $\chi$ as a function of $T$ for various system sizes $L$;  (b) The Binder cumulant and (c) correlation length normalized by the size $L$ as a function of $T$ for increasing $L$;  [$J=0$, and $V=4J'$ for (a)-(c)].}
\label{fig3}
\end{figure*}

\section{Finite temperature phase diagram: a thermodynamic phase transition in a 1D dissipative interacting quantum system}
The most striking effect of the SB coupling on the quantum many-body system can be found at finite temperature. On the one hand,  it is well known that there is no spontaneous symmetry breaking at any finite temperature for a closed 1D system  with local Hamiltonian (while long-range interactions may change this scenario~\cite{Thouless1969,Dyson1969,Kosterlitz1976,Imbrie1988,Spohn1989,Luijten2001}). The argument goes as follows, for instance for the CDW phase:  Thermal fluctuations induce pairs of kink-antikink domain walls which cost only  a finite energy. At any finite temperature, the entropy gain by  deconfining the excitations overwhelms the energy cost. The leads to a proliferation of domain walls, which destroys the long-range CDW order.
Furthermore, quantum fluctuations induced by the SB coupling are also detrimental to the CDW order. On the other hand, the infinite compressibility of the DW phase may alter the domain-wall proliferation argument.
%At the same time, the quantum fluctuations induced by the SB coupling are also detrimental to the CDW order, thus one may wonder  what will happen in the presence of  both the thermal and quantum fluctuations simultaneously.

In order to better understand the latter, we switch off the quantum fluctuations in the system ($J=0$), and calculate the DW (staggered) susceptibility of the system, defined as $\chi= \frac \beta L (\langle m^2\rangle -\langle |m|\rangle^2)$\cite{Kotze2008}, with $m=\frac 1L\sum_i (-1)^i n_i$. We plot $\chi$ as a function of $T$ in  Fig.~\ref{fig3} (a). Other than the upturn associated with the ground state previously discussed, we observe a peak which keeps growing with system size and whose peak position extrapolates to a finite value of temperature in the thermodynamic limit (see the inset of Fig.~\ref{fig3} (a) for the finite size scaling of the peak position), which is a signature of a finite temperature phase transition. To verify this point, we also calculate the Binder cumulant $U_2$ and the correlation length $\xi_a$ defined as
\begin{eqnarray}
U_2=\frac 32 (1-\frac 13 \frac{\langle m^4\rangle}{\langle m^2\rangle^2})\\
\xi_a=\frac 1{q_1}\sqrt{\frac{S(\pi)}{S(\pi+q_1)}-1}
\end{eqnarray}
where $S(Q)=1/L^2 \sum_{ij} e^{iQ(i-j)}\langle(n_i-\frac 12)(n_j-\frac 12)\rangle$  and $q_1=2\pi/L$.  The $U_2$ and $\xi_a/L$ as a functions of $T$ with different system sizes are plotted in Fig.~\ref{fig3} (b) and (c), both of which have exhibited the signature of phase transition.

 However, it is difficult to infer the universality class of the transition from the numerical data: the lowest system sizes in Fig.~\ref{fig3} (a) are apparently not in the scaling regime and the values of $\chi$ attained are rather low. Due to the presence of the bath we expect the finite size effects to be more important than for the usual spin or bosonic models with one particle. The finite size effects can also been seen in Fig.~\ref{fig3} (b) and (c), where we don't find an strictly size independent (common crossing) point. To study the universal class of this phase transition, we need to simulate significantly larger system within the scaling regime. This is difficult in our current QMC simulations with worm-type update due to the special comb-like geometry of the system lattice, especially in the case we studied with $J=0$, where the update may be inefficient even though the model is free from sign problem. Also, due to the limited system sizes we can study and the strong finite size effects, we can't preclude the possibility of an crossover, where the position of peak in Fig.~\ref{fig3} (a) may shift extremely slowly to zero with increasing system size.

\begin{figure}[htb]
\includegraphics[width=0.485\linewidth,bb=76 59 547 530]{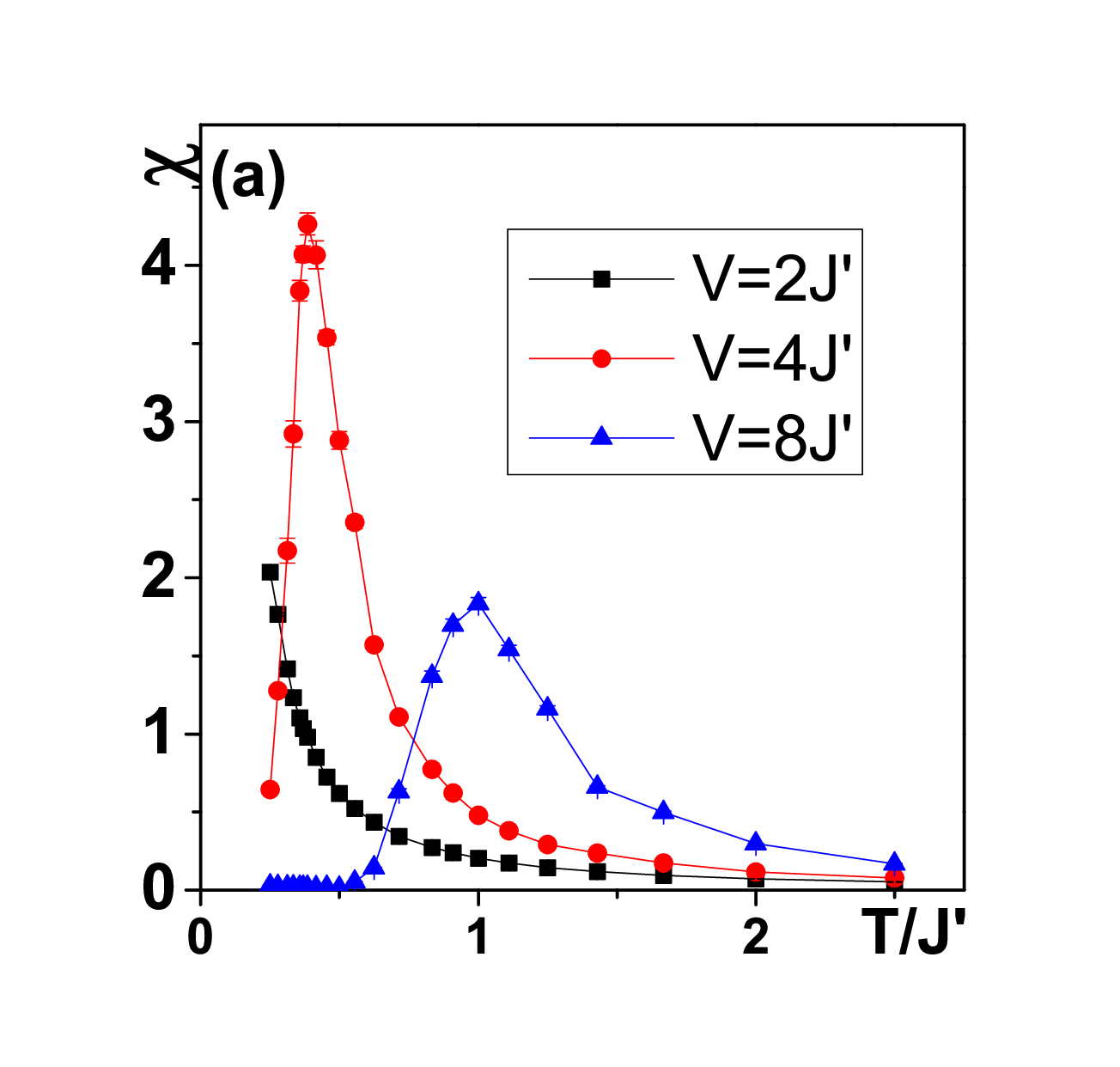}
\includegraphics[width=0.495\linewidth,bb=56 59 547 536]{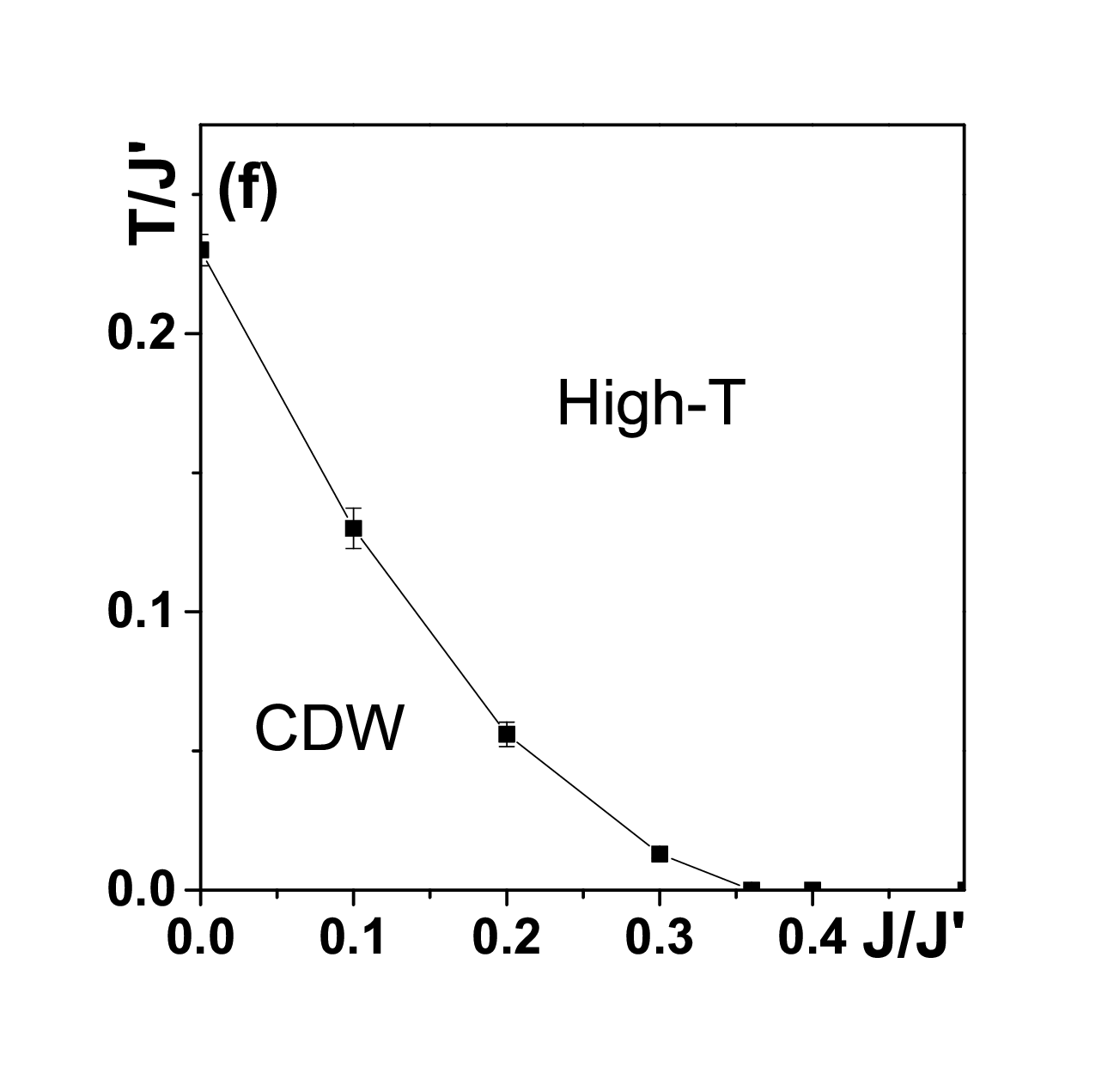}
\caption{ (a) $\chi$ as a function of $T$ with $L=64$, $J=0$ and different $V$; (b)the finite temperature phase diagram in the $T-J$ plane with $V=4J'$.}
\label{fig4}
\end{figure}

If we start from a system whose ground state has no DW order ({\it e.g.} $V=2J'$) and increase the temperature, there is no finite temperature phase transition, as shown in Fig.~\ref{fig4} (a). Hence, even though thermal fluctuations and the SB coupling are both detrimental to the DW long-range order, they can conspire to facilitate it. Similar phenomena have observed before in the finite temperature phase transition of a quantum compass model in a square lattice with $\sigma^z_i\sigma^z_{i+e_x}$ coupling along the horizontal bonds and $\sigma^x_i\sigma^x_{i+e_y}$ coupling along the vertical bonds playing a similar role of quantum fluctuations\cite{Wenzel2008}. The reason behind this is that the total system we studied has a comb-like geometry in 2D. The domain wall excitations are no longer deconfined: due to the SB coupling, separating a kink-antikink pair over some distance $d$ will inevitably perturb the bath chains between them, and cost an energy depending on $d$. In the case of models with long range interactions~\cite{Thouless1969,Dyson1969,Kosterlitz1976,Imbrie1988,Spohn1989,Luijten2001}), the interactions between a kink-antikink pair are also confined, which originates from the long-range nature of the interactions, instead of the system-bath couplings as in our case.      In the previous discussion in this section, we set $J=0$ thus all the quantum fluctuations come from the off-diagonal SB coupling terms. To complete our discussion, we switch on the single particle hopping in the system chain, and study the finite temperature phase diagram of the model in the $T-J$ plane with a fixed $V=4J'$, the result is shown in the Fig.~\ref{fig4} (b). \\

\section{Discussion}
While it is widely believed that dissipation leads to dephasing which drives a quantum system towards a classical one, here we show an example that the quantum dissipation induced by an off-diagonal SB coupling to a quantum environment, can enhance the quantum fluctuations in the system and give rise to counterintuitive phenomena.    One of the most important questions is to what extent real dissipation can be modeled  by the specific choice of the bath and SB coupling used in this paper, which provides the simplest example of gapless quantum modes coupling locally and off-diagonally to the system. Usually, the randomness feature of the environment can be absorbed into the spectrum function of the bath, while a proper revision of the bath Hamiltonian in our model can mimic more general dissipation with various spectral functions. However, our model doesn't apply for those open quantum many-body systems with non-local dissipation where the entire system shares the same environment (as {\it e.g.} in cavity QED\cite{Cecilia2013}). Also, the open quantum systems coupled to a classical environment {\it e.g.} a classical noise involves the real time evolutions\cite{Poletti2012,Cai2013,Sieberer2013,Marino2016}, thus are beyond the scope of our current scheme.

\section{ Conclusion and Outlook}
We propose a conceptually simple approach to deal with quantum dissipation in open interacting quantum systems, which allows us to investigate both Markovian and non-Markovian physics ranging from weak to strong SB coupling regimes in a unified picture. We explored the properties of an embedded 1D  quantum many-body system by numerically solving a special comb-like lattice geometry. Counterintuitive zero-temperature and thermal behavior has been discovered, indicating that dissipation can fundamentally alter the way we look at the properties of quantum many-body systems: spontaneous symmetry breaking can occur in a subsystem of reduced dimensionality spatially embedded in a larger system with an inhomogeneous Hamiltonian.
Some avenues for further work immediately suggest themselves. First, a generalization to higher dimensions is straightforward. Notice that both the bath and the SB coupling term lead to positive definite expansions in the path integral formulation. Therefore, as long as the system Hamiltonian is free from the sign problem, the total Hamiltonian can be solved in a numerically exact way  by QMC. This allows one in principle to investigate the effect of quantum dissipation on various systems with long-range order (symmetry breaking) and even topological order, but also on how quantum coherence and entanglement properties may change. Even though in this paper we only study the equilibrium state (for the total system), it would be very interesting to explore the non-equilibrium dynamics of the dissipative quantum many-body system in the current scheme, which is beyond the scope of QMC simulations but may be accessible by other methods (e.g. an exactly solvable quadratic fermionic model defined on a similar comb-like lattice). The far-from-equilibrium dynamical and steady state properties of a driven-dissipative quantum many-body system remain an open and elusive question\cite{Hartmann2016}.

\appendix

\begin{figure}[htb]
\includegraphics[width=0.7\linewidth,bb=1 1 200 187]{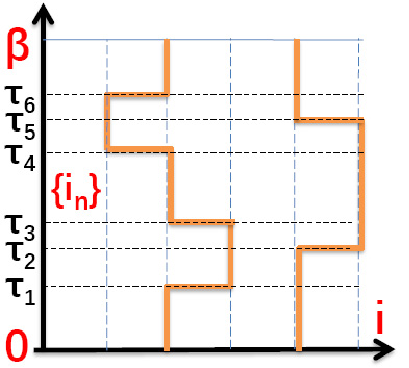}
\caption{A typical space-(imaginary) time configurations in quantum Monte Carlo simulations.}
\label{fig:config}
\end{figure}

\section{A Brief introduction of the quantum Monte Carlo method with worm algorithm}

We consider a Hamiltonian $\hat{H}$ which can be decomposed as $\hat{H}=-\hat{T}+\hat{V}$ , where the $\hat{T}$ represents the off-diagonal terms in the Hamiltonian and $V$ is the diagonal ones in the Fock basis $|i\rangle=|n_1n_2\cdots n_L\rangle$. The partition function $Z=\mathbf{Tr}e^{-\beta\hat{H}}$ can be expanded in terms of the probability functions of space-imaginary time configurations as
\begin{eqnarray}
\nonumber Z&=&\mathbf{Tr}\sum_{n=0}^\infty\int_0^\beta
d\tau_n\int_0^{\tau_n} d\tau_{n-1}\cdot\cdot\cdot \int_0^{\tau_2}
d\tau_1\\
\nonumber  &\times&e^{-\tau_1\hat{V}}\hat{T}e^{-(\tau_2-\tau_1)\hat{V}}\cdots
e^{-(\tau_n-\tau_{n-1})\hat{V}}\hat{T}e^{-(\beta-\tau_n)\hat{V}}\\
\nonumber &=&\sum_{n=0}^\infty \sum_{|i_1\rangle,\cdots|i_n\rangle}
\int_0^\beta d\tau_n\int_0^{\tau_n} d\tau_{n-1}\cdot\cdot\cdot
\int_0^{\tau_2} d\tau_1\\
&\times& W(\tau_1,\cdots,\tau_n,|i_1\rangle,\cdots,|i_n\rangle) \label{eq:partition}
\end{eqnarray}
where $\tau_n$ is the time of the $n$th tunneling event in the world-line of the particles, and $|i_m\rangle$ denotes the particle configurations between the
imaginary time $\tau_{m-1}$ and $\tau_m$ (as shown in Fig.\ref{fig:config}), and for a given space-imaginary time configuration: $\{\tau_1,\cdots,\tau_n,|i_1\rangle,\cdots,|i_n\rangle\}$, the corresponding probability $W(\tau_1,\cdots,\tau_n,|i_1\rangle,\cdots,|i_n\rangle)$ can be written as:
\begin{eqnarray}
\nonumber W=\langle i_1|\hat{T}|i_2\rangle e^{-(\tau_2-\tau_1)E_{i_2}}\langle
i_2|\hat{T}|i_3\rangle e^{-(\tau_3-\tau_2)E_{i_3}}\cdots\\
\times e^{-(\tau_n-\tau_{n-1})E_{i_n}}\langle i_n|\hat{T}|i_1\rangle
e^{-(\beta+\tau_1-\tau_n)E_{i_1}}
\end{eqnarray}
where $E_{i_m}=\langle i_m|\hat{V}|i_m\rangle$ is the interaction (diagonal) energy for particle
configuration between $\tau_m$ and $\tau_{m-1}$. As long as $\hat{T}$ is a positive definite operator, we can prove that $W(\tau_1,\cdots,\tau_n,|i_1\rangle,\cdots,|i_n\rangle)$ is always positive, which enables us to perform the importance sampling and evaluate the average value of physical quantities over a limited number of space-(imaginary) time configurations. A worm algorithm is a update algorithm where the partition function in Eq.(\ref{eq:partition}) is sampled indirectly  in the extended configuration space of open world-line configurations by performing local movement\cite{Prokofev1998}.

\section{Physical quantities obtained by quantum Monte Carlo (QMC)}

In this section, we will discuss the derivations of several physical quantities from the QMC simulations. We should emphasize that all these quantities are only defined for the system instead of the system+bath.

\subsection{Superfluid density}

The superfluid density $\rho_s$ is defined as

\begin{equation}
\rho_s=\frac {L}{\beta} \langle W^2\rangle,
\end{equation}
where $W$ is the winding number along the system chain,  which can be obtained in the quantum Monte Carlo (QMC) simulations\cite{Ceperley1987}. Considering a world line of  a hard-core boson in the imaginary time-space configurations, even though a system boson can escape into the bath from certain system site, it will finally return to the system via the same system site, since each bath chain  is independent  and a world line  should be a closed curve.

\subsection{Compressibility}

The system compressibility $\kappa$ is defined as the system particle number $N=\sum_i\langle n_i\rangle $ in response to the perturbation $H'=-\sum_i \mu n_i$, which only operates in the system sites:
\begin{equation}
\kappa=\frac 1L\frac {d}{d\mu} N(\mu)|_{\mu=0}
\end{equation}
We assume that the bosons in the bath chain don't feel the chemical potential. As we demonstrated in the main text, the bosons in the bath chains don't necessarily mean the real bosonic particles as in the system, they can be any bosonic energy reservoir, {\it e.g.} phonon, photon and so on,  which is insensitive to the external fields  operating on the system bosons. We can prove that $\kappa=\frac {\beta}{L} \Delta N$,
where $\beta=1/T$ is the inverse temperature, $L$ the system size, and $\Delta N=\langle
\hat{N}^2\rangle-\langle \hat{N}\rangle^2$ the variance of the system particle
number.  Notice that
\begin{equation}
N(\mu)=\frac 1 {Z(\mu)} Tr (\sum_i \hat{n}_i) e^{-\beta \hat{H}_{tot}+\beta\mu\sum_i \hat{n}_i}
\end{equation}
with $Z(\mu)= Tr e^{-\beta \hat{H}_{tot}+\beta\mu\sum_i \hat{n}_i}$, thus
\begin{eqnarray}
\nonumber \frac {d N(\mu)}{d\mu}&=&\frac \beta {Z(\mu)} Tr \hat{N}^2 e^{-\beta \hat{H}_{tot}+\beta\mu\sum_i \hat{n}_i}\\
\nonumber &-&\frac 1 {Z^2(\mu)} \frac{\partial Z(\mu)}{\partial\mu}Tr \hat{N} e^{-\beta \hat{H}_{tot}+\beta\mu\hat{N}}\\
&=&\beta (\langle N^2\rangle - \langle N\rangle^2)
\end{eqnarray}
In the numerical simulation, the particle number fluctuations can be derived
from the histogram of the system particle number distributions in the QMC
simulations, which is assumed to be a Gaussian distribution for sufficient large system. We fit the distribution of the particle number
by a Gaussian distribution as $P(N)=\frac
{1}{\sqrt{2\pi\sigma}}e^{-\frac{(N-\bar{N})^2}{2\sigma}}$ and the system particle number fluctuation is the width of the continuous Gaussian
distribution (in the thermodynamic limit), $\Delta N=\sigma$.

\begin{figure}[htb]
\includegraphics[width=0.8\linewidth,bb=19 19 282 210]{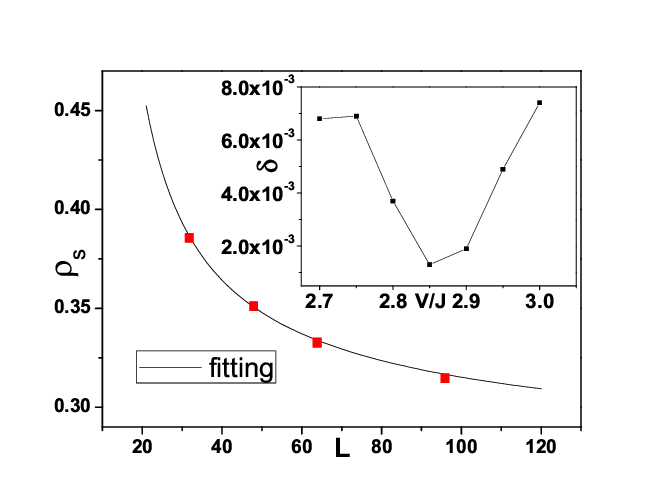}
\caption{Analysis of the KT phase transition for $J'=0.2J$ and $V=2.85J$ using the Weber-Minnhagen fitting form of Eq(\ref{eq:sflow}). The average error of the fitting is shown in the inset as a
function of $V$ for fixed $J'=0.2J$. }
\label{fig:S}
\end{figure}

\subsection{Charge-density-wave susceptibility}

The charge-density-wave (CDW) susceptibility of the system is defined as the static linear susceptibility of the CDW order parameter $m=\frac 1L \langle \sum (-1)^i n_i\rangle$ in response to the perturbation $H'=-\mu_s\sum_i (-1)^i n_i$ only operating on the system sites :
\begin{equation}
\chi=\frac {d}{d\mu_s} m(\mu_s)|_{\mu_s=0}=\frac LT (\langle m^2\rangle -\langle m\rangle^2)
\end{equation}
For any finite system size $\langle m\rangle=0$, thus in the QMC simulation, we use a revised CDW susceptibility $\chi'$ defined as:
\begin{equation}
\chi'=\frac LT (\langle m^2\rangle -\langle |m|\rangle^2)
\end{equation}
which agrees with $\chi$ the in thermodynamic limit.

\section{The Koserlitz-Thouless transition} \label{appendix:K}
At the Kosterlitz-Thouless (KT) phase transition point, the superfluid density flows as a
function of $L$ as~\cite{Weber1988}:
\begin{equation}
\rho_s^c(L)=\rho_s^c(L\rightarrow \infty)(1+\frac {1}{2\ln(L)+C}), \label{eq:sflow}
\end{equation}
where $\rho_s^c(L\rightarrow \infty)$ is the critical value of the superfluid density  at the
transition point in the thermodynamic limit, and  $C$ is a non-universal constant.  To determine the $\rho_s^c$ and the critical point $V_c$, we use the Weber-Minnhagen fit of Eq(\ref{eq:sflow}) to obtain the superfluid density calculated in our QMC simulations,
$\rho_s^{\rm QMC}(L)$, and calculate the average error of the fit,
$\delta=\frac 14\sum_L |\rho_s^{\rm fit} (L)-\rho_s^{\rm QMC}(L)|$ with $L=32,48,64,96$, for different interaction strengths $V$. Since the flow equation Eq.(\ref{eq:sflow}) is only valid at the phase transition point, we expect that the the average error will reach it minimum at the critical point $V=V_c$, as shown Fig.\ref{fig:S}.

{\it Acknowledgements -- }  ZC wishes to thank P. Zoller for fruitful discussions.
ZC acknowledges the support from the National Key Research and Development Program of China (grant No. 2016YFA0302001), the National Natural Science Foundation of China under Grant No.11674221 and the Shanghai Rising-Star Program.  LP is supported by FP7/ERC starting grant No. 306897 and acknowledges funding by DFG through NIM-2. X. Wang was supported by the National Program on Key Research Project (Grants No. 2016YFA0300501), and by the National Natural Science Foundation of China (Grant No. 11574200). ZY, JL and YC are supported by the National Natural Science Foundation of China (Grant No. 11625416 and 11474064) and the State Key Programs of China (Grant No. 2016YFA0300504). We acknowledge the support from the Center for High Performance Computing of Shanghai Jiao Tong University.

%\bibliography{real}

\end{document}